\documentclass[showpacs,aps,prc,nofootinbib,showkeys,unsortedaddress,raggedbottom,twocolumn]{revtex4-1}
\pdfoutput=1

\usepackage{bm}
\usepackage{amsmath}
\usepackage{amssymb}
\usepackage{graphicx}
\usepackage{subfigure}
\usepackage{float}
\usepackage[usenames,dvipsnames]{color}
\definecolor{darkblue}{RGB}{0,0,196}
\usepackage[colorlinks=true,linkcolor=darkblue,citecolor=darkblue,urlcolor=darkblue]{hyperref}

\usepackage{setspace}
\usepackage{footmisc}
\usepackage[makeroom]{cancel}

\newcommand{\intdP}{\int\!dP}

\def\be{\begin{equation}}
\def\ee{\end{equation}}
\def\ba{\begin{eqnarray}}
\def\ea{\end{eqnarray}}

\begin{document}

\title{Kaonic Hanbury-Brown-Twiss radii at 200 GeV and 5.02 TeV}

\author{Mubarak Alqahtani} 
\address{Department of Physics, College of Science, Imam Abdulrahman Bin Faisal University, Dammam 31441, Saudi Arabia}

\author{Michael Strickland} 
\affiliation{Department of Physics, Kent State University, Kent, OH 44242 United States}

\begin{abstract}
We use 3+1D quasiparticle anisotropic hydrodynamics (aHydroQP) to make predictions for kaon Hanbury-Brown-Twiss (HBT) radii in 200 GeV and 5.02 TeV heavy-ion collisions.  Using previously determined aHydroQP parameters, we compute kaonic HBT radii and their ratios as a function of the mean transverse momentum of the pair $k_T$.  We first consider Au-Au collisions at 200 GeV, finding good agreement between aHydroQP predictions and experimental data up to $k_T \sim 0.8$ GeV. We then present predictions for kaonic HBT radii and their ratios in 5.02 TeV Pb-Pb collisions. Our aHydroQP predictions do not exhibit a clear $k_T$ scaling of the pion and kaon source radii, however, an approximate transverse mass $m_T$ scaling is observed, particularly at 200 GeV.
\end{abstract}

\date{\today}


\maketitle


A quark-gluon plasma (QGP) is expected to be created in ultrarelativistic heavy-ion collisions at the Relativistic Heavy Ion Collider (RHIC) at Brookhaven National Laboratory and the Large Hadron Collider (LHC) at CERN. This occurs because such collisions create a region of high energy density in which quarks and gluons are deconfined for a short time.  
The generated matter hydrodynamizes on a time scale of roughly 1 fm/c and the initial temperatures generated have been found to well exceed the temperature of the QGP crossover transition, $T_\text{QGP} \sim 155$~MeV~\cite{Florkowski:2010zz,Chaudhuri:2012yt,Borsanyi:2013bia,HotQCD:2014kol,Haque:2014rua,Averbeck:2015jja,Braun-Munzinger:2015hba,Busza:2018rrf,Almaalol:2020rnu}. 

Comparison of a variety of experimental observables, such as the identified particle spectra and elliptic flow with predictions from relativistic viscous hydrodynamics have indicated that dissipative hydrodynamics can be used to understand experimental observations in a variety of collision systems.  To describe this collective behavior, different dissipative hydrodynamic frameworks have been developed such as second-order viscous hydrodynamics and anisotropic hydrodynamics, see e.g., Refs.~\cite{Romatschke:2009im,Niemi:2011ix,Heinz:2013th,Gale:2013da,Strickland:2014pga,Shen:2014vra,Ryu:2015vwa,Jaiswal:2016hex,Jeon:2016uym,Romatschke:2017ejr,Florkowski:2017olj,Romatschke:2017ejr,Alqahtani:2017jwl,Giacalone:2017dud,Sievert:2019zjr}.

In this work, we use 3+1D quasiparticle anisotropic hydrodynamics (3+1D aHydroQP) \cite{Alqahtani:2015qja} to study kaonic Hanbury-Brown-Twiss (HBT) radii.  The 3+1D aHydroQP includes the relevant ingredients necessary to compute HBT interferometry. First, in the hydrodynamic stage, the 3+1D dynamical equations are non-conformal and use a realistic lattice-based equation of state.  The dynamics includes both shear and bulk viscous effects in addition to an infinite number self-consistently determined higher-order transport coefficients \cite{Alalawi:2020zbx}. In the next stage, the 3+1d aHydroQP evolution is converted to hadrons using an anisotropic Cooper-Frye freezeout, which is implemented using a customized version of THERMINATOR 2 \cite{1102.0273}. This package includes both hadronic production and  decays.  The customized version samples from anisotropic hadron distributions, which are guaranteed to be non-negative even far from equilibrium~\cite{Alqahtani:2016rth}. 

HBT correlations are quantum mechanical in nature and can provide information about the space-time structure of the source; its size, shape, and the emission duration.  The 3+1D quasiparticle anisotropic framework has been previously applied to compute pionic HBT radii at both LHC and RHIC energies, finding quite good agreement between theoretical predictions and experimental observations~\cite{Alqahtani:2017tnq,Alqahtani:2020daq}

We first present predictions for kaonic HBT radii in $\sqrt{s_{NN}}$ = 200 GeV Au-Au collisions. We find good agreement between 3+1D aHydroQP predictions and experimental results available from PHENIX collaboration. Next, we present predictions for kaonic HBT radii in $\sqrt{s_{NN}}$ = 5.02 TeV Pb-Pb collisions, where thus far no experimental data have been reported. We present the $k_T$-dependence of the HBT radii and their ratios in different centrality classes. 

Next, we study the transverse mass $m_T = \sqrt{k_T^2 + m^2}$ dependence in order to understand the particle-species dependence of the HBT radii. We find that the $k_T$ scaling of pion and kaon source radii is broken, while an approximate transverse-mass $m_T$ scaling is observed, particularly at 200 GeV. The existence of $m_T$ ($k_T$) scaling would imply that the measured HBT radii for particles with different masses, such as pions and kaons, would collapse onto a single curve when plotted versus $m_T$ $(k_T)$.

Our paper is organized as follows.  In Sec.~\ref{sec:model}, the 3+1D quasiparticle anisotropic hydrodynamics model is introduced.  In Sec.~\ref{sec:results}, we compare our findings at 200 GeV with experimental results for kaonic HBT radii. Next, we present predictions for kaonic HBT radii in 5.02 TeV collisions. Finally, Sec.~\ref{sec:conclusions} contains our conclusions and an outlook for the future.

\section{Model}
\label{sec:model}

In this section, we will briefly describe the main ingredients of quasiparticle anisotropic hydrodynamics, the freeze-out method, and the hadronic afterburner employed.  The hydrodynamic runs used were tuned to identified-particle spectra in prior papers \cite{Almaalol:2018gjh,Alqahtani:2020paa} and full details of the methods employed can be found in Refs.~\cite{Alqahtani:2015qja,Alqahtani:2016rth,Alqahtani:2017jwl,Alqahtani:2017tnq} and two reviews \cite{Alqahtani:2017mhy,Alalawi:2021jwn}. 

Firstly, the evolution of the hydrodynamic stage is governed by 3+1D dissipative dynamical equations, which are obtained from the 1st and 2nd moments of the Boltzmann equation in relaxation time approximation~\cite{Alqahtani:2015qja,Alqahtani:2017mhy,Alalawi:2021jwn}. For quasiparticles having thermal masses $m(T)$, the Boltzmann equation is given by~\cite{Jeon:1995zm,Berges:2005md,Romatschke:2011qp,Alqahtani:2015qja}
\be
p^\mu \partial_\mu f(x,p)+\frac{1}{2}\partial_i m^2\partial^i_{(p)} f(x,p)=-C[f(x,p)] \, ,
\label{eq:boltzmanneq}
\ee
where $C[f(x,p)]$ is the collisional kernel which accounts for all (non mean-field) interactions. Here, we take the collisional kernel to be given in the relaxation-time approximation.

The first moment of Eq.~\eqref{eq:boltzmanneq}, which corresponds to the requirement of conservation of energy and momentum, gives 
\be
\partial_\mu  T^{\mu \nu} = 0\, \label{eq:1stmoment} ,
\ee
where
\be
T^{\mu\nu} = \int dP \, p^\mu p^\nu  f \, ,
\ee
while the second moment of Eq.~\eqref{eq:boltzmanneq} gives
\be
\partial_\alpha  I^{\alpha\nu\lambda}- J^{(\nu} \partial^{\lambda)} m^2 =-\intdP \, p^\nu p^\lambda{\cal C}[f]\, \label{eq:2ndmoment} ,
\ee
where the particle four-current is defined by
\be
J^\mu = \int dP \, p^\mu f \, ,
\ee
and $I^{\mu\nu\lambda}$ is defined by
\be
I^{\mu\nu\lambda} = \int dP \, p^\mu p^\nu p^\lambda f \, ,
\ee
with the Lorentz invariant integration measure being $\int dP = N_\text{dof} \int \frac{d^3{\bf p}}{(2\pi)^3} \frac{1}{E}$.  Parentheses around Lorentz indices indicates symmetrization, e.g., $A^{(\mu\nu)} = (A^{\mu\nu}+A^{\nu\mu})/2$

In the aHydroQP approach, a realistic equation of state obtained from lattice quantum chromodynamics (QCD) calculations is used to determine the temperature dependence of the mass \cite{Alqahtani:2015qja}.   The standard implementation used herein is based on the equation of state extracted by the Wuppertal-Budapest lattice QCD collaboration \cite{Borsanyi:2010cj}.  

The underlying one-particle distribution function is assumed to be anisotropic in momentum space 
\be
f(x,p) =  f_{\rm eq}\!\left(\frac{1}{\lambda}\sqrt{\sum_i \frac{p_i^2}{\alpha_i^2} + m^2}\right) .
\label{eq:fform}
\ee
with only diagonal momemtum-space anisotropies encoded in $\alpha_i$ with $i = 1,2,3$. We note that $\alpha_i(t,{\bf x})$ and the temperature-like scale $\lambda(t,{\bf x})$ are functions of space and time. 

The system is initialized using smooth Glauber initial conditions assuming Bjorken flow in the longitudinal direction and zero transverse flow ($u_x(\tau_0)=u_y(\tau_0)=0$). For RHIC 200 GeV Au-Au collisions, the initial temperature at $\tau_0 = 0.25$ fm/c is taken to be $T_0 = 455$ MeV and the specific shear viscosity is  assumed to be $\eta/s =  0.179$, which were found to best reproduce experimental results in Ref.~\cite{Almaalol:2018gjh}.
For LHC 5 TeV Pb-Pb collisions, the initial temperature at $\tau_0 = 0.25$ fm/c is taken to be $T_0 = 630$ MeV  and the specific shear viscosity is  assumed to be $\eta/s = 0.159$, which again were found to best reproduces experimental results in Ref.~\cite{Alqahtani:2020paa}. We note that having fixed the specific shear viscosity, the bulk viscosity and all other higher-order transport coefficients are fixed uniquely.  Further details concerning the derivation of the dynamical equations and initialization of the system in the aHydroQP approach can be found in Refs.~\cite{Nopoush:2014pfa,Alqahtani:2015qja,Alqahtani:2016rth,Alqahtani:2017mhy,Almaalol:2018gjh}. 

The second phase of the evolution starts after the system reaches the freeze-out temperature.  Based on prior works, we take $T_{\rm FO}=130$ MeV, see e.g.~\cite{Alqahtani:2017jwl,Alqahtani:2017tnq,Almaalol:2018gjh,Alqahtani:2020paa}. Using an anisotropic Cooper-Frye prescription, the number of hadrons produced on a constant energy-density hypersurface is calculated from the underyling space-time dependent hydrodynamic variables such as the flow velocity and anisotropy parameters~\cite{Alqahtani:2017mhy}. In this phase, a customized version of THERMINATOR 2~\cite{1102.0273} is used to perform the statistical production and decays of hadrons.  Both the aHydroQP and THERMINATOR 2a codes are publicly available online using Ref.~\cite{MikeCodeDB}.

\subsection{Interferometry}
\label{sec:HBT}

In this section, we will review the correlation functions used in the analysis of HBT radii.  These correlations exist between pairs of identical particles due to their quantum mechanical nature. Experimentally, a two-particle correlation $C$ is constructed as the ratio $C(q) = A(q)/B(q)$ with $q$ being the relative momentum given by $q=\vec{p}_{2}-\vec{p}_{1}$. $A(q)$ is the measured momentum difference distribution between two particles with $p_1$ and $p_2$ taken from the same event (actual pairs), while $B(q)$ is a reference distribution of pairs of particles picked randomly from different events in the same centrality bin (mixed pairs), see e.g. \cite{nucl-th/9901094,1504.05168,ALICE:2017iga}.  To compare to experiment, we compute these correlation functions using event generator models which perform statistical hadronization.

The identical particle two-particle correlation function is defined as
\be
C(\vec{p}_{1},\vec{p}_{2})=\frac{W_{2}(\vec{p}_{1},\vec{p}_{2})}{W_{1}(\vec{p}_{1})W_{1}(\vec{p}_{2})} \, ,
\label{eq:C}
\ee
where $W_{1}$ is a one-particle distribution and $W_{2}$ is the two-particle distribution function. Both distributions can be obtained by the space-time integral of a Wigner phase-space density (or the emission function) $S(x,p)$, which encodes the probability of emission of a particle with momentum $p$ from a space-time point $x$~\cite{nucl-th/9901094} 
\begin{equation}
W_{1}(\vec{p})= E_{p}\frac{dN}{d^3p} \int d^4x \, S(x,p) \, ,
\label{eq:W1}
\end{equation}
and
\begin{eqnarray}
W_{2}(\vec{p}_{1},\vec{p}_{2}) &=& E_{p1}E_{p2}\frac{dN}{d^3p_1d^3p_2}
\nonumber \\
 && \times \int \! \! \int d^4x_1 \, d^4x_2 \, S(x_1,x_2,p_1,p_2)  \, . \;\;
\label{eq:W2}
\end{eqnarray}
We note that in the absence of quantum correlations, $C(\vec{p}_{1},\vec{p}_{2})$ would be unity. Assuming a source with a Gaussian profile, the emission function can be parameterized as
\be
S(r)=  N \, {\rm exp}\!\left(-\frac{r^{2}_{{\rm out}}}{2R^{2}_{{\rm out}}}-\frac{r^{2}_{{\rm side}}}{2R^{2}_{{\rm side}}}-\frac{r^{2}_{{\rm long}}}{2R^{2}_{{\rm long}}} \right) \, ,
\label{eq:S}
\ee
with $N$ being a normalization constant, $r_i^2$ are the squared relative space-time separation of the pair, and $R_i^2$ are the squared HBT radii. In this formula, the subscript $i$ represents different directions: out (parallel to the pair's $k_T$), long (parallel to the beam axis), and side (perpendicular to both long and out).

 In this analysis, we will use the Bertsch-Pratt parameterization~\cite{Pratt:1986cc,Bertsch:1989vn}, which is performed in the Longitudinal Center-of-Mass System (LCMS) where the mean longitudinal momentum of the pair vanishes. Using the Bertsch-Pratt parameterization and Eq.~\eqref{eq:S},  the two-particle correlation function can be written in terms of three Gaussians.
\be
 C(q,k)=  1+\lambda \, e^{-R^{2}_{{\rm out}}q^{2}_{{\rm out}}-R^{2}_{{\rm side}}q^{2}_{{\rm side}}-R^{2}_{{\rm long}}q^{2}_{{\rm long}}} \, .
\label{eq:fitform}
\ee
with $q$ being the relative momenta decomposed as well into three components, $\vec{q}=$(${q_{\rm out}}$, ${q_{\rm side}}$, ${q_{\rm long}}$). $\lambda$ is the normalization factor (also called the incoherence parameter) which characterizes the correlation strength. For core-halo systems, it can deviate from unity, $0 \leq \lambda \leq 1$  where $\lambda=1$ for chaotic sources and $\lambda=0$ for totally coherent sources~\cite{Akkelin:2001nd}. It is also affected by long-lived resonances \cite{Wiedemann:1996ig} and non-Gaussian features of $S(r)$ \cite{Bystersky:2005qx}.  For squeezed states it is possible to have $\lambda > 1$~\cite{Asakawa:1998cx,Csorgo:2007iv,Dudek:2010nr}.  In our analysis, $\lambda$ is taken as a free fit parameter.  We note that Coulomb repulsion between similar charge pairs is ignored in the parameterization specified in Eq.~\eqref{eq:fitform}. This is consistent with the fact that experimental results have been corrected for these effects. The HBT radii are obtained by fitting to the correlation function as a function of $k_T$.  The HBT radii provide information about the (in-)homogeneity of the system (its effective sizes). In general, $R_{ij}$ has mixed spatial and temporal information of the source. For example, the product $R_{\rm out} \, R_{\rm side} \, R_{\rm long}$ gives the effective volume and $R_{\rm out}^2$ is sensitive to the emission time. We refer the reader to Refs.~\cite{nucl-ex/0505014,nucl-th/9901094,Florkowski:2010zz,Chaudhuri:2012yt}, for more information about the relation between the HBT radii and the space-time structure of the final freeze-out stage. 

 We note that long-lived resonances and hadronic elastic rescattering are not included in our
simulations using THERMINATOR 2. Both effects can result in non-Gaussian modifications of the underlying correlation functions ~\cite{Shapoval:2014wya,Kincses:2019rug}. Without such effects included, it is reasonable to assume that the correlations functions have a Gaussian form. We also note that such approximations are frequently used in modeling results,  e.g., see Refs.~\cite{Chakraborty:2020tym,Plumberg:2020jod}.  In a forthcoming paper we plan to include these effects.  In the meantime, we refer the reader to recent studies that examined different effects on the shape of the HBT correlation functions~\cite{Cimerman:2020tpd,Kincses:2022eqq}.

\begin{figure*}[t!]
\centerline{
\hspace{-1.5mm}
\includegraphics[width=0.9\linewidth]{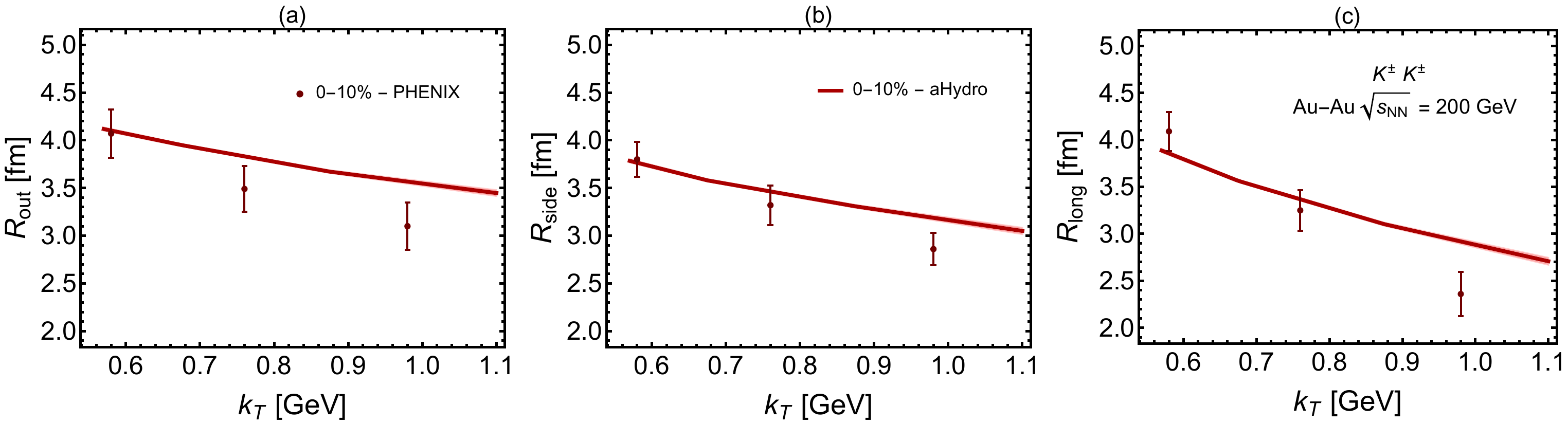}}
\centerline{
\hspace{-1.5mm}
\includegraphics[width=0.9\linewidth]{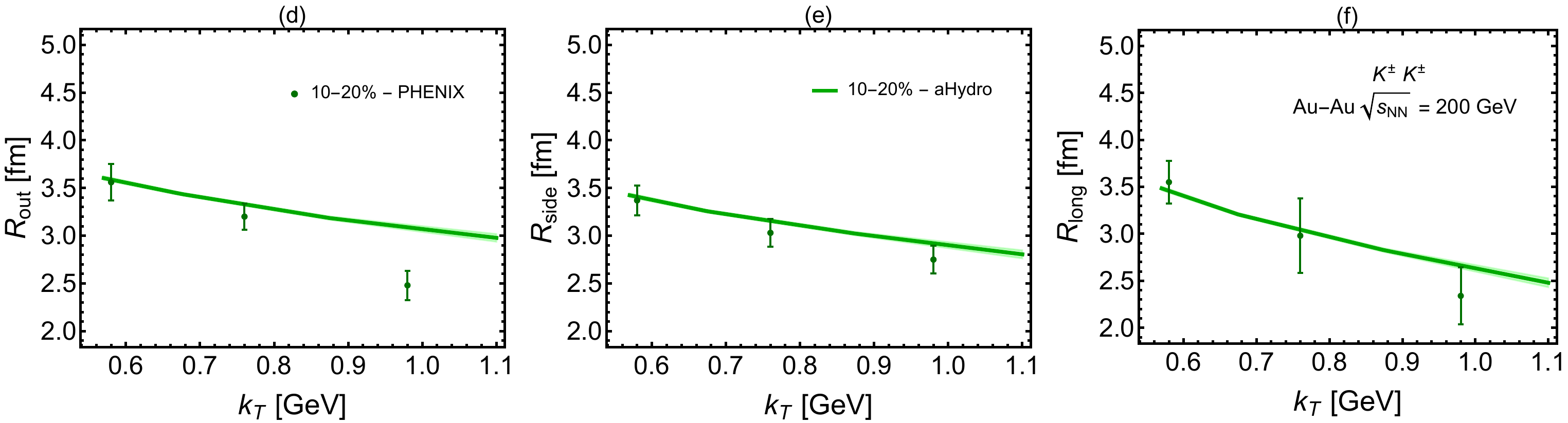}}
\caption{The $k_T$ dependence of kaonic HBT radii in the 0-10$\%$ (top row) and 10-20$\%$ (bottom row) centrality classes.  The extracted $R_{\rm out}$, $R_{\rm side}$, and $R_{\rm long}$ radii are shown in the left, middle, and right columns, respectively. Data are from PHENIX collaboration at 200 GeV~\cite{1504.05168}.}
\label{fig:HBTradiiPH}
\end{figure*}

\begin{figure*}[t!]
\centerline{
\hspace{-1.5mm}
\includegraphics[width=0.9\linewidth]{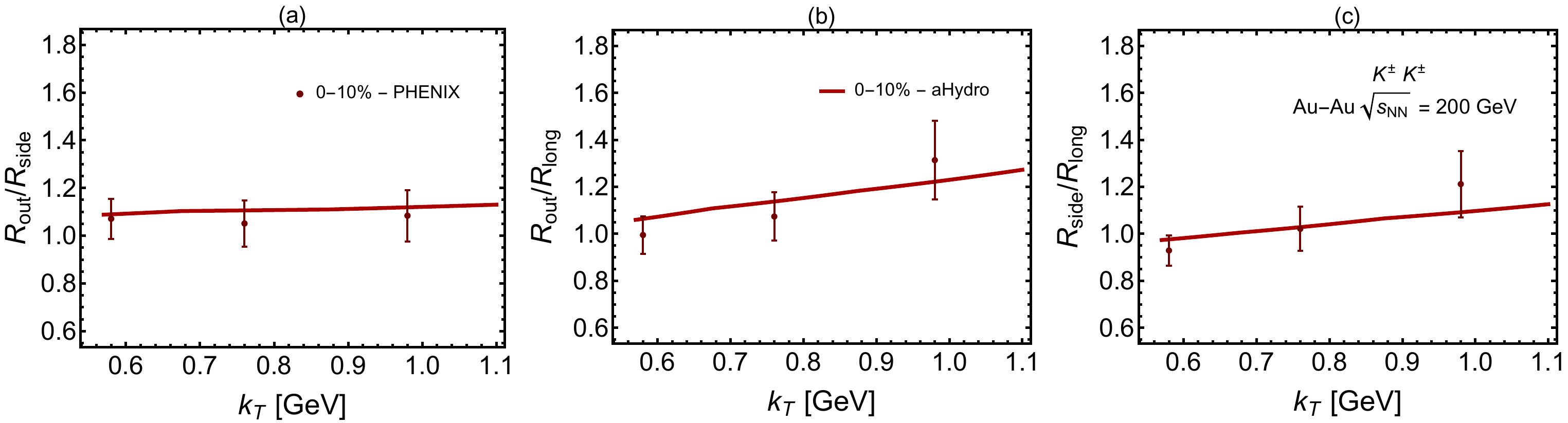}}
\centerline{
\hspace{-1.5mm}
\includegraphics[width=0.9\linewidth]{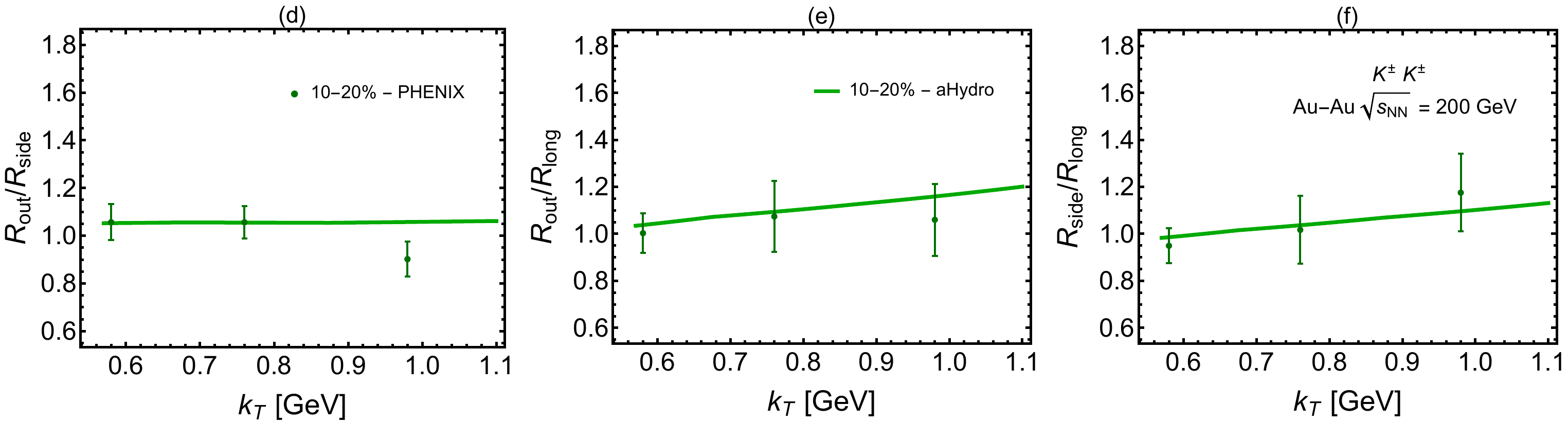}}
\caption{The $k_T$ dependence of the ratios of kaonic HBT radii for charged kaons in the 0-10$\%$ (top row) and 10-20$\%$ (bottom row) centrality classes.  The ratios $R_{\rm out}/R_{\rm side}$, $R_{\rm out}/R_{\rm long}$, and $R_{\rm side}/R_{\rm long}$ are shown in the left, middle, and right columns, respectively. Data are from PHENIX collaboration at 200 GeV~\cite{1504.05168}. }
\label{fig:HBTratiosPH}
\end{figure*}

\begin{figure*}[t!]
\centerline{
\hspace{-1.5mm}
\includegraphics[width=0.9\linewidth]{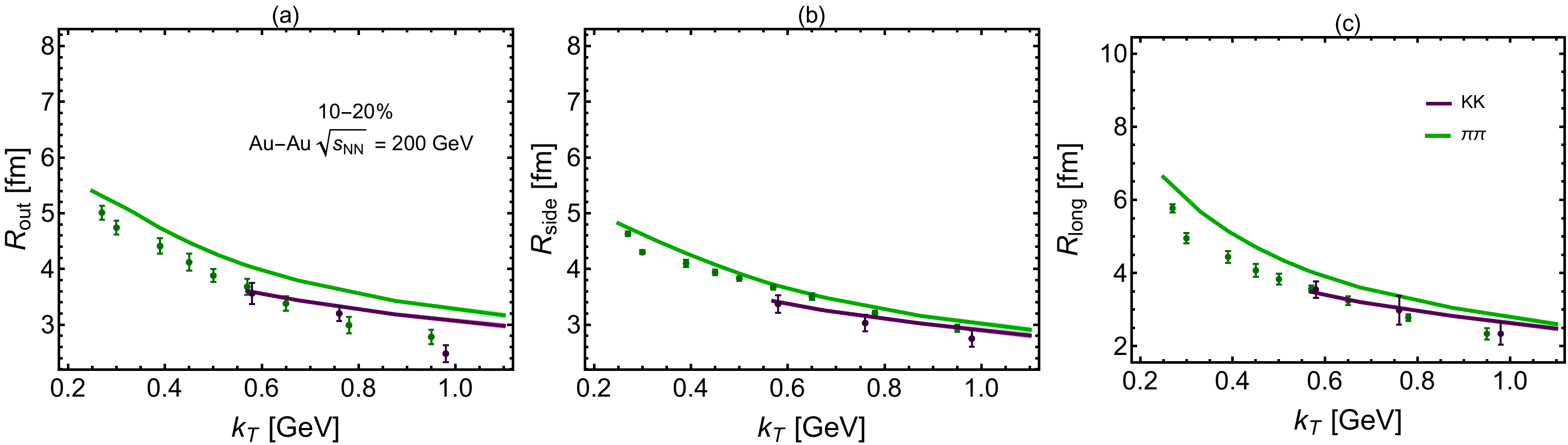}}
\centerline{
\hspace{-1.5mm}
\includegraphics[width=0.9\linewidth]{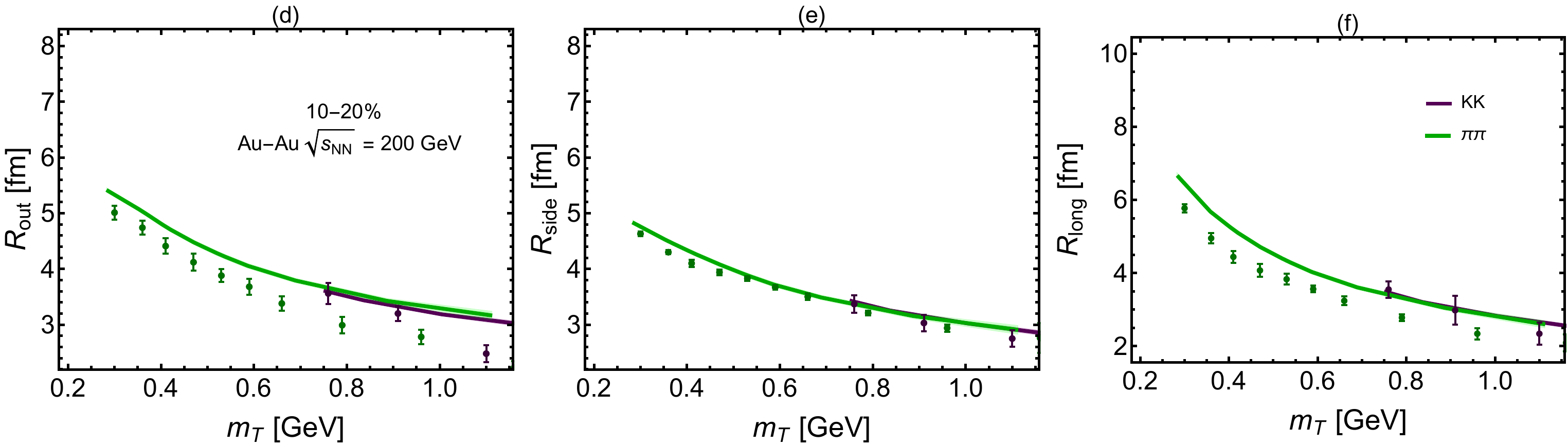}}
\caption{HBT radii as a function of $k_T$ (top row) and $m_T$ (bottom row) for charged pions \cite{Alqahtani:2020daq}  and kaons  in the 10-20$\%$ centrality class.  The extracted $R_{\rm out}$, $R_{\rm side}$, and $R_{\rm long}$ radii are shown in the left, middle, and right columns, respectively. Data are from PHENIX collaboration at 200 GeV: for pions~\cite{PHENIX:2004yan} whereas for kaons~\cite{1504.05168}.}
\label{fig:HBTradiiPH-mt}
\end{figure*}

\begin{figure*}[t!]
\centerline{
\hspace{-1.5mm}
\includegraphics[width=0.95\linewidth]{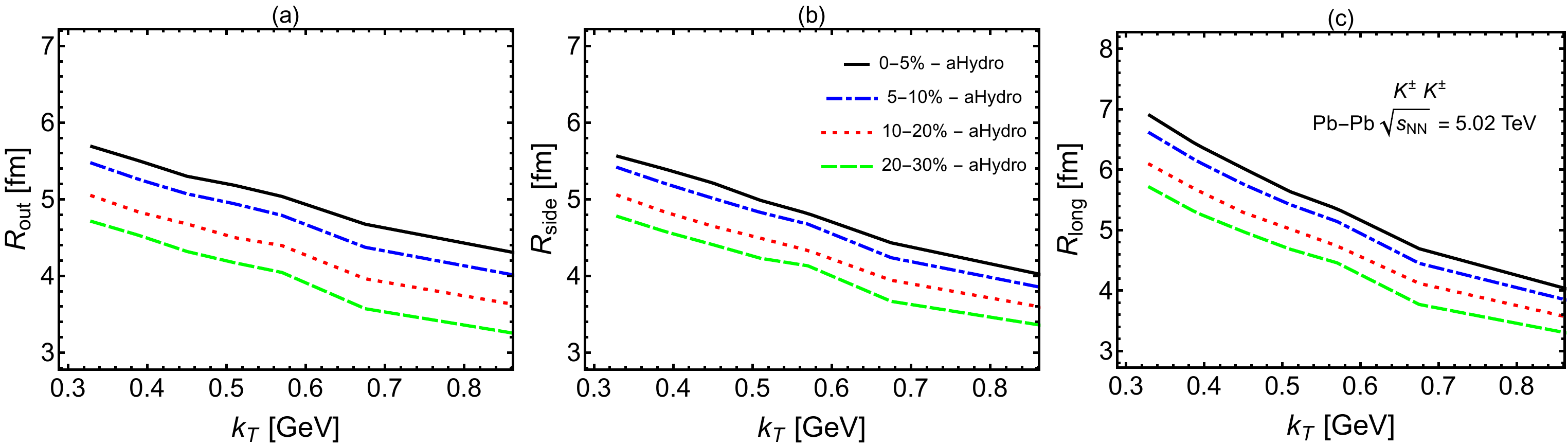}}
\caption{The $k_T$ dependence of the kaonic HBT radii in the 0-5$\%$, 5-10$\%$, 10-20$\%$, and 20-30$\%$ centrality classes.  The extracted $R_{\rm out}$, $R_{\rm side}$, and $R_{\rm long}$ radii are shown in the left, middle, and right columns, respectively. All results are predictions of 3+1D aHydroQP model for 5.02 TeV Pb-Pb collisions. Statistical uncertainty bands are not shown, however, they are listed in Table.~\ref{table:HBTradiidata} and are on the order of $\sim 1$ \%. }
\label{fig:HBTradii5TeV}
\end{figure*}

\begin{figure*}[t!]
\centerline{
\hspace{-1.5mm}
\includegraphics[width=0.95\linewidth]{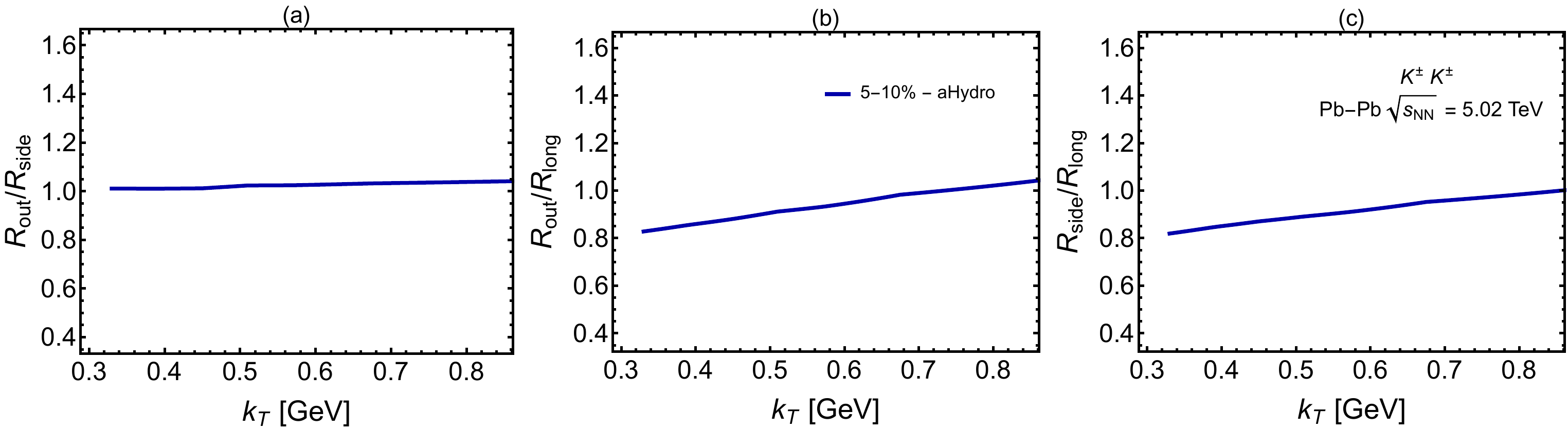}}
\centerline{
\hspace{-1.5mm}
\includegraphics[width=0.95\linewidth]{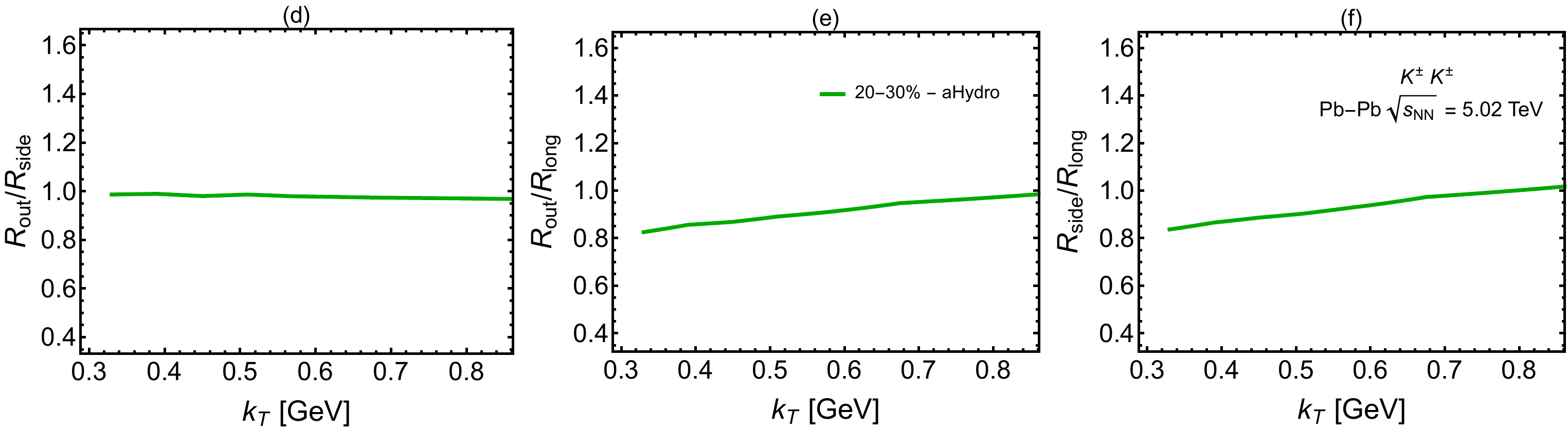}}
\caption{Ratios of the kaonic HBT radii as a function of $k_T$ in the 5-10$\%$, and 20-30$\%$ centrality classes.  The ratios $R_{\rm out}/R_{\rm side}$, $R_{\rm out}/R_{\rm long}$, and $R_{\rm side}/R_{\rm long}$ are shown in the left, middle, and right columns, respectively. All results are predictions of 3+1D aHydroQP model for 5.02 TeV Pb-Pb collisions.}
\label{fig:HBTratios5TeV}
\end{figure*}

\begin{figure*}[t!]
\centerline{
\hspace{-1.5mm}
\includegraphics[width=0.95\linewidth]{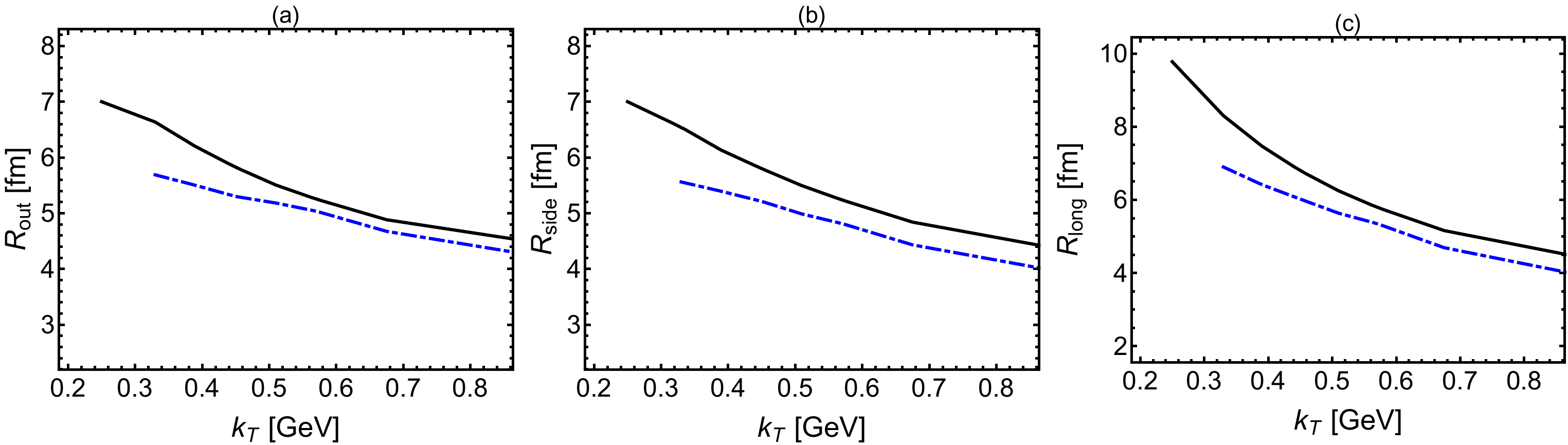}}
\centerline{
\hspace{-1.5mm}
\includegraphics[width=0.95\linewidth]{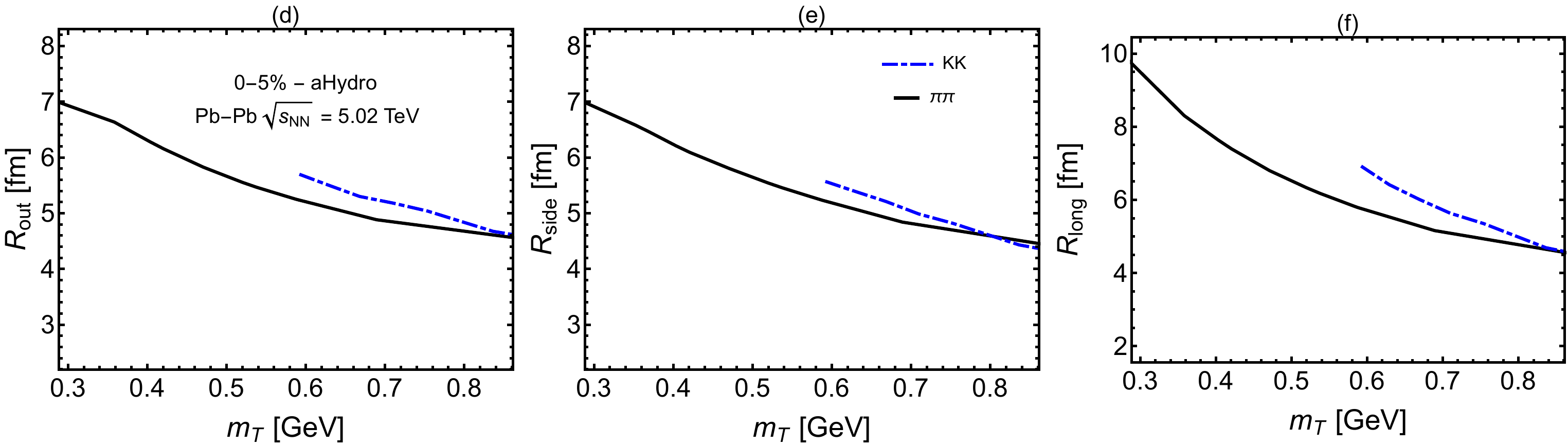}}
\caption{HBT correlation radii as a function of $k_T$ (top row) and $m_T$ (bottom row) for charged pions \cite{Alqahtani:2020paa} (black solid lines) and kaons (blue dashed lines) in the 0-5$\%$ centrality class.  $R_{\rm out}$, $ R_{\rm side} $, and $R_{\rm long}$ are shown in the left, middle, and right columns, respectively. All results are predictions of 3+1D aHydroQP model for 5.02 TeV Pb-Pb collisions.}
\label{fig:HBTradii5TeVpiK}
\end{figure*}

\begin{figure*}[t!]
\centerline{
\hspace{-1.5mm}
\includegraphics[width=0.67\linewidth]{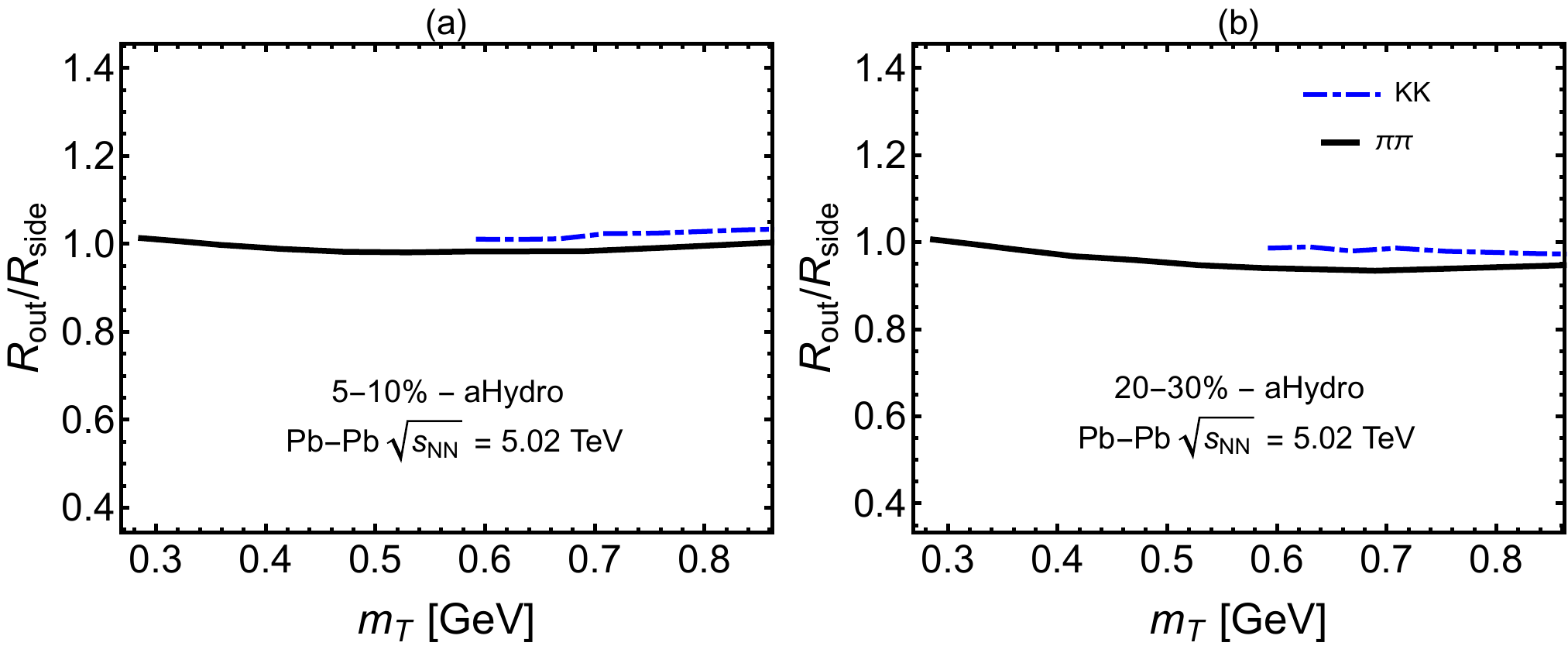}}
\caption{$R_{\rm out}/R_{\rm side} $ as a function of $m_T$ for charged pions \cite{Alqahtani:2020paa} (black solid lines) and charged kaons (blue dashed lines) in the 5-10$\%$ and  20-30$\%$ centrality classes, left and right panels, respectively. All results are predictions of 3+1D aHydroQP model for 5.02 TeV Pb-Pb collisions.}
\label{fig:HBTratios5TeVpiK}
\end{figure*}

\section{Results}
\label{sec:results}

Before presenting our predictions for the kaonic HBT radii in 5.02 TeV Pb-Pb collisions, we first present comparisons with PHENIX data collected in 200 GeV Au-Au collisions in order to gauge the accuracy of our predictions. In Figs.~\ref{fig:HBTradiiPH} and \ref{fig:HBTratiosPH} we consider the transverse momentum range  $0.3 < p_T < 1.5$ GeV in order to match the experimental cuts used in Ref.~\cite{0903.4863}. We use the following relative transverse momentum, $k_T$, bins $0.54{-}0.6$, $0.6{-}0.75$, $0.75{-}1.0$, and $1.0{-}1.2$ GeV and plot the values obtained at the center of each of these bins. With this choice of bins, we cover the experimental bins which had the following means, $k_T \in \{0.58, 0.76, 0.98\}$~GeV.  For values in between our bin centers, we use linear interpolation to interpolate our results. Note that in the 10-20\% centrality class we do not show the experimental point with the highest $k_T$ since the aHydro prediction differs significantly from the experimental result at high $k_T$. 

In Fig.~\ref{fig:HBTradiiPH}, we show the $k_T$-dependence of the kaonic HBT radii $R_{\rm out}$, $R_{\rm side} $, and $R_{\rm long}$ in the 0-10\% and 10-20\% centrality classes in the top and bottom rows, respectively. In both centrality classes, we see that our model agrees quite well with PHENIX experimental data up to $k_T \sim 0.8$ GeV.
At larger $k_T$ the aHydroQP predictions are above the experimental data. Next, in Fig.~\ref{fig:HBTratiosPH} we present the $k_T$-dependence of the HBT radii ratios, $R_{\rm out}/R_{\rm side}$, $ R_{\rm out}/R_{\rm long} $, and $R_{\rm side}/R_{\rm long}$, in the same centrality classes.  Again the agreement of aHydroQP with PHENIX data is quite good up to $k_T \sim 1.0$ GeV in both centrality classes. We note that, at the same collision energy and small $k_T$, one finds similar agreement between aHydroQP predictions and experimental data for the pionic HBT radii~\cite{2007.04209}. We note that the agreement between theory and data reported here is much better, particularly at low $k_T$.

We next investigate the $k_T$ and transverse mass $m_T$ scaling of the HBT radii where $m_T=\sqrt{k_T^2+m^2}$ with $m$ being the particle's mass (kaon or pion).
Fig.~\ref{fig:HBTradiiPH-mt} shows the extracted HBT radii of charged pions and kaons in the 10-20\% centrality class as a function of $k_T$ (top row) and $m_T$ (bottom row). As can be seen from the top row, the data and also our model results do not show any sign of $k_T$ scalling. On the other hand, from the bottom row, our results indicate that there is scaling at high $m_T$ since, within
our uncertainties, the pion and kaon predictions match smoothly onto one another. There
does seem to be a lack of strict scaling in the experimental data. We note, however, that both our results and the data show that the $R_{\rm side}$ shows an approximate scaling~\cite{1504.05168}. 
 
Next, we present our predictions for the kaonic HBT radii and their ratios in 5.02 TeV Pb-Pb collisions.  In Figs.~\ref{fig:HBTradii5TeV} and \ref{fig:HBTratios5TeV}, to match the experimental cut used by the ALICE collaboration at 2.76 TeV~\cite{ALICE:2017iga}, we consider the transverse momentum range  $0.15 < p_T < 1.5$ GeV. In the case of 5.02 TeV collisions, the following $k_T$ bins were considered for the charged kaons $0.3{-}0.36$, $0.36{-}0.42$, $0.42{-}0.48$, $0.48{-}0.54$, $0.54{-}0.60$, $0.60{-}0.75$, and $0.75{-}1.0$ GeV. We then linearly interpolate our results as described above.

In Fig.~\ref{fig:HBTradii5TeV}, we present the kaonic HBT radii, $R_{\rm out}$, $ R_{\rm side} $, and $R_{\rm long}$, as a function of $k_T$ in the 0-5\%, 5-10\%, 10-20\%, and 20-30\% centrality classes. As can be seen from this figure, the radii are smaller for more peripheral collisions than for central ones. The radii also decrease with increasing $k_T$ due to the collective flow of the QGP. Next, in Fig.~\ref{fig:HBTratios5TeV}, we show comparisons of the kaonic HBT radii ratios, $R_{\rm out}/R_{\rm side} $, $ R_{\rm out}/R_{\rm long} $, and $R_{\rm side}/R_{\rm long}$, in the 5-10\%, and 20-30\% centrality classes. We see that the ratio $R_{\rm out}/R_{\rm side}$ is consistently close to unity in all centrality classes considered here. We note that, in the case of pions, experimental measurements of $R_{\rm out}/R_{\rm side}$ indicate values that are slightly less than unity at higher $k_T$, see Fig.~4 of Ref.~\cite{Alqahtani:2020daq}.

We next investigate the $k_T$ and transverse mass $m_T$ scaling of the HBT radii at 5.02 TeV collisions. To test whether the $k_T$ and $m_T$ scaling hold, we compare results of this work with the pionic HBT radii obtained in Ref.~\cite{Alqahtani:2020paa} using the same model, 3+1D aHydroQP. From the top panel of Fig.~\ref{fig:HBTradii5TeVpiK}, we see that the radii do not show a clear scaling with the $k_T$. The $m_T$ dependence of the HBT radii are shown in the bottom panel of Fig.~\ref{fig:HBTradii5TeVpiK}. One can see that there is only evidence of $m_T$-scaling at high $m_T$.  At low $m_T$ both the kaonic $R_{\rm out}$ and $R_{\rm long}$ are predicted to be larger than the corresponding pionic HBT radii at the same transverse mass. In Fig.~\ref{fig:HBTradii5TeVpiK}, we only show the $k_T$ and $m_T$ dependence of the HBT radii in the 0-5$\%$ centrality class, however, a similar behaviour is seen in all centrality classes considered in this work. In Fig.~\ref{fig:HBTratios5TeVpiK}, the $m_T$ dependence of  $R_{\rm out}/R_{\rm side} $ is shown in 5-10\% and 20-30\% centrality classes. As can be seen in both panels, the ratio 
$R_{\rm out}/R_{\rm side} $ is predicted to be slightly larger for kaons than for pions, which indicates different space-time correlations. Similar findings were reported in Ref.~\cite{ALICE:2017iga} by the ALICE collaboration in 2.76 TeV Pb-Pb collisions. The breaking of the $m_T$ scaling, particularly for $R_{\rm long}$ has also been observed by the PHENIX collaboration~\cite{1504.05168}. The violation
of $m_T$-scaling has been suggested to be a result of rescatterings which differently influence pions and kaons, see Refs.~\cite{Shapoval:2014wya,Sinyukov:2015kga}, however, herein, we do not include such rescatterings and still find scaling violation. We also note that using a framework of (3+1)D viscous hydrodynamics combined with
THERMINATOR 2, the authors of~\cite{Kisiel:2014upa,2010.12161} found an approximate $m_T$ scaling of the HBT radii at 2.76 TeV and 5.02 TeV, respectively. To the best of our knowledge, the kaonic femtoscopy experimental results at 5.02 TeV are not yet available to compare our model predictions against, so our predictions now wait to be compared with experiment.

\section{Conclusions and outlook}
\label{sec:conclusions}

In this paper, we used 3+1D quasiparticle anisotropic hydrodynamics to analyze charged kaon femtoscopy. The aHydroQP hydrodynamical parameters were tuned in separate studies and herein we simply applied those to make predictions for kaonic HBT radii.  To treat the post-hydrodynamic hadronic evolution we made use of a customized version of the THERMINATOR 2 package. 

We first presented comparisons with data in Au-Au collisions at 200 GeV. Comparisons of the $k_T$ dependence of the HBT radii showed a good agreement between aHydroQP and experimental results at low $k_T$, however, for $k_T \gtrsim 0.8$ GeV we observed differences between the predictions of aHydroQP and the experimental results, which may be due to the use of smooth initial conditions for our hydrodynamic simulations. We also found that the data and also our model results do not show any sign of $k_T$ scaling. On the other hand, our results indicate that there is scaling at high $m_T$, whereas the data itself does not show a clear scaling except for $R_{\rm side}$, which shows an approximate scaling.

Next we presented aHydroQP predictions for the three-dimensional femtoscopic radii 
in Pb–Pb collisions at 5.02 TeV nucleon-nucleon collision energy. All results show a decrease of the source radii $R_i$  with increasing transverse $k_T$ and decreasing multiplicity (increasing centrality). Our model predictions also do not show a clear $k_T$ or $m_T$-scaling of pion and kaon source radii in the $k_T$ and $m_T$ ranges considered in this study; however, there are perhaps indications of $m_T$ scaling at high $m_T$.  

Finally we note that the existence of $m_T$ scaling of the HBT radii was based on early ideal hydrodynamic models ~\cite{Beker:1994qv,Chapman:1994yv,Akkelin:1995gh,Csorgo:1995bi,NA44:2001hbu}. It is interesting that at RHIC energies this scaling emerges even when dissipative corrections are included.  It may even hold at high $m_T$ at LHC energies; however, herein we observed clear violations of $m_T$ scaling at low $m_T$ for 5.02 TeV collisions.

Looking to the future, we plan to include the effect of fluctuating initial conditions in aHydroQP and the effect of using the URQMD and SMASH afterburners \cite{Bass:1998ca,Bleicher:1999xi,Weil:2016zrk}.

\section*{Acknowledgments}

We would like to thank Pritam Chakraborty for fruitful discussions regarding \cite{2010.12161} and THERMINATOR 2. M. Strickland was supported by the U.S. Department of Energy, Office of Science, Office of Nuclear Physics under Award No. DE-SC0013470.

\appendix 

\begin{widetext}

\section{}
\label{app:data}

For reference, in Table.~\ref{table:HBTradiidata}, the results of 3+1D aHydroQP model for all HBT radii are listed. The $k_T$ bins considered here are:  $0.3{-}0.36$, $0.36{-}0.42$, $0.42{-}0.48$, $0.48{-}0.54$, $0.54{-}0.60$, $0.60{-}0.75$, and $0.75{-}1.0$~GeV. 
%
\begin{table*}[h!]
\centerline{
\hspace{-1.5mm}
\includegraphics[width=0.8\linewidth]{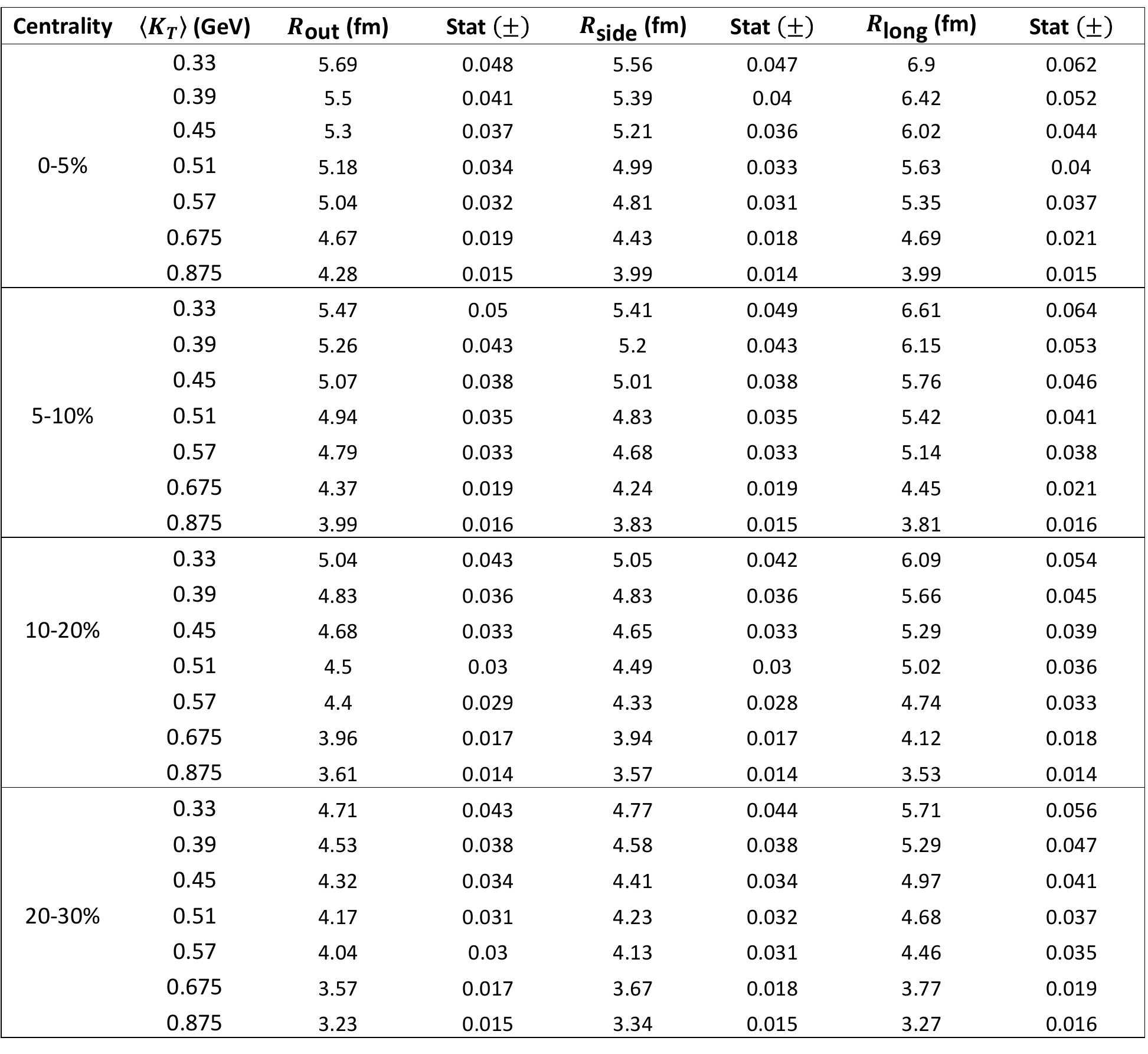}}
\caption{The $k_T$ dependence of kaonic HBT radii in four centrality classes. We note that the statistical uncertainties are absolute values in femtometers. All results are predictions of the 3+1D aHydroQP model for 5.02 TeV Pb-Pb collisions.}
\label{table:HBTradiidata}
\end{table*}

\end{widetext}

\newpage

\bibliographystyle{apsrev4-1}
\bibliography{KaonsHBT}

\end{document}